\magnification=1095
\raggedbottom
\baselineskip=16truept
\overfullrule 0pt
\raggedbottom
\footline={\ifnum\pageno>1 \hfil -- \folio\ -- \hfil \else\hfil\fi}
\headline={}

\def\sec{$^{\prime\prime}$} 
\def\deg{$^\circ$}

\def\hugeskip{\vskip 0.5 truein}

 at 10truept
 at 10truept
\font\lbf cmbx12 at 16truept

\input psfig.tex

\font\bigsf=cmssbx10 scaled 1400
\hbox{
\psfig{figure=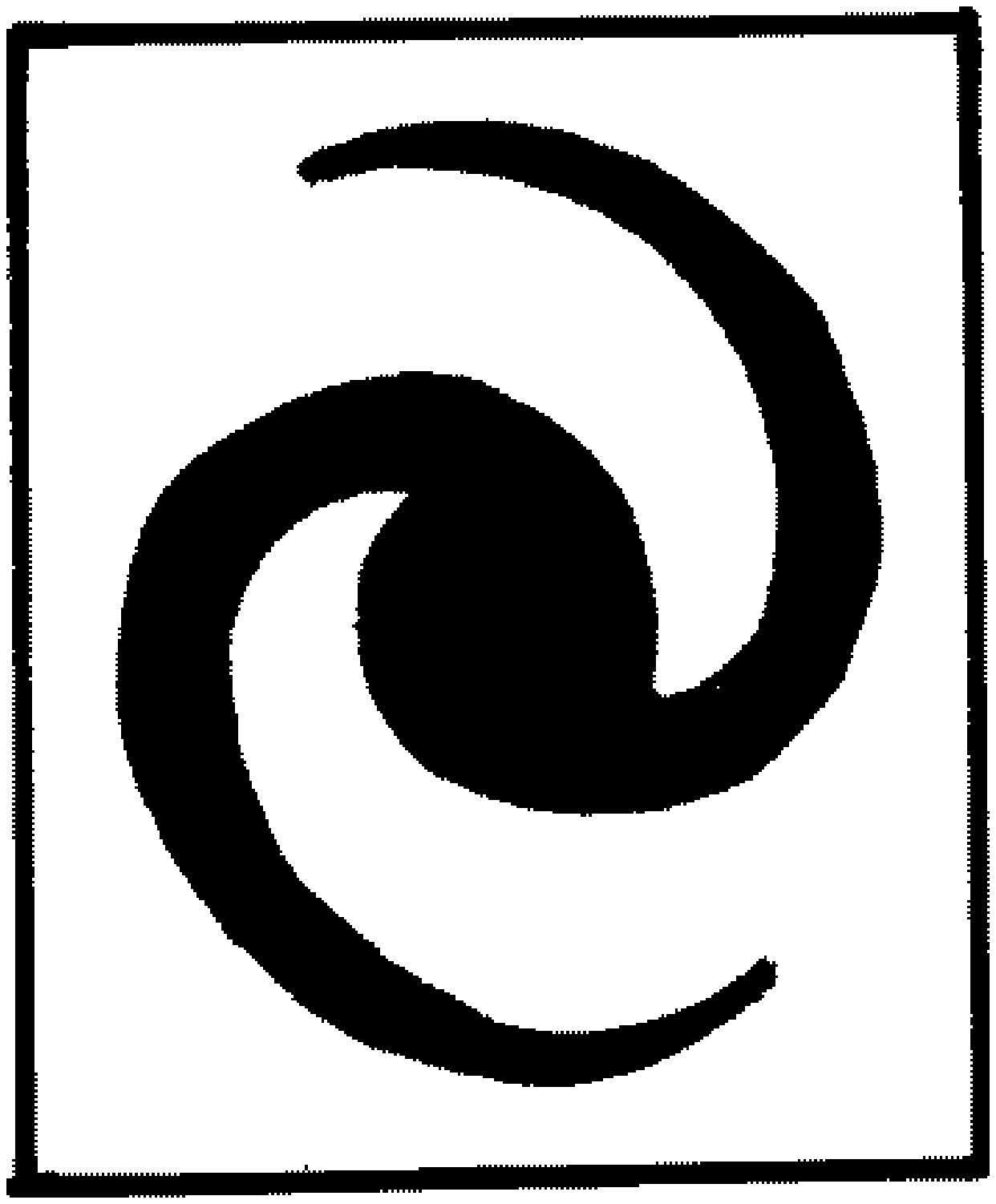,height=1in,width=1in}
\hskip 0.5truein 
\vbox to 1. truein {\bigsf \vfill
\noindent UNIVERSITY OF CALIFORNIA AT BERKELEY 
\medskip \noindent ASTRONOMY DEPARTMENT
\vfill}
}

\hugeskip \hugeskip

\centerline{\lbf EVIDENCE FOR A PRECESSING ACCRETION DISK} 
\bigskip
\centerline{\lbf IN THE NUCLEUS OF NGC~1097}

\hugeskip \noindent
\centerline{Thaisa Storchi-Bergmann$^{\; 1}$, Michael Eracleous$^{\; 2,3}$, 
Maria Teresa Ruiz$^{\; 4,5}$,}
\centerline{Mario Livio$^{\; 6}$, Andrew S. Wilson$^{\; 7}$, and
Alexei V. Filippenko$^{\; 3}$}

\hugeskip
\centerline{To Appear in the {\it Astrophysical Journal}}
\hugeskip

\footnote{}{
\item{$^1$}
Instituto de F\'\i sica, UFRGS, CP 15051, Porto Alegre, RS, Brasil,
\hfill\break e-mail: {\tt thaisa@if.ufrgs.br}.
\item{$^2$}
Hubble Fellow.
\item{$^3$}
Department of Astronomy, University of California, Berkeley, CA 94720-3411,
\hfill\break e-mail: {\tt mce@beast.berkeley.edu}, 
{\tt alex@wormhole.berkeley.edu}.
\item{$^4$}
Universidad de Chile, Casilla 36-D, Santiago, Chile,
e-mail: {\tt mtruiz@das.uchile.cl}.
\item{$^5$}
Visiting Astronomer, Cerro Tololo Inter-American Observatory. 
CTIO is operated by AURA, Inc.\ under contract to the National Science
Foundation.
\item{$^6$}
Space Telescope Science Institute, 3700 San Martin Drive, Baltimore, MD 21218,
USA, \hfill\break e-mail: {\tt mlivio@stsci.edu}.
\item{$^7$}
Astronomy Department, University of Maryland, College Park, MD 20742,
\hfill\break e-mail: {\tt wilson@astro.umd.edu} .
\smallskip}

\vfill \eject

\centerline{\vbox{\hsize=5.5 truein
\centerline{\bf Abstract}
\medskip 
\noindent
We present new spectroscopic observations of the LINER (and now
Seyfert~1) nucleus of NGC~1097, and discuss the evolution of its
broad, double-peaked Balmer lines. When originally discovered in 1991,
the red peak of the double-peaked H$\alpha$ line was stronger than the
blue, while by 1994 the H$\alpha$ profile had become almost symmetric
and the integrated line flux had decreased to half its original
value. Our new spectrum, taken in 1996, shows that the broad,
double-peaked lines have returned to almost their original strengths,
the profiles of H$\beta$ and H$\alpha$ are identical to within errors,
and the broad-line emitting region is unreddened. However, the profile
of the Balmer lines is now such that the blue peak is stronger than
the red, opposite to the asymmetry observed in 1991. Various
models are considered for the observed behavior, all assuming that the
emission lines originate in an accretion disk. We present a refined
version of the precessing, planar, elliptical accretion ring model
proposed by Storchi-Bergmann et al. and Eracleous et al. This model
provides an acceptable fit to the line profiles. We also consider the
possibility that the line profile evolution results from a precessing
warp in the disk, induced by irradiation from the center, and show
that the range of radii and precession time scales expected in this
model are consistent with the observations. The sudden appearance of
the ``disk-like'' broad line profiles in NGC~1097 could have resulted
from the formation of a new accretion disk due to, for example, the
tidal disruption of a star, or the illumination of a pre-existing disk 
by a transient ionizing source at the center of the disk.
\medskip\noindent
{\it Subject Headings} galaxies: individual (NGC~1097)--galaxies: nuclei--
galaxies: Seyfert -- accretion disks -- line:profiles
}}

\bigskip

\centerline{\bf 1. Introduction}
\medskip

In the most widely accepted picture for active galactic nuclei (AGNs),
a nuclear, supermassive black hole is fed by an accretion
disk. Nevertheless, observational evidence for the presence of the
disk has been limited and indirect until very recently (see, for
example, Koratkar et al. 1992, 1995; Antonucci et al. 1989,
1992). There was no clear, dynamical signature of the disk in the form
of double-peaked emission lines, such as those found in cataclysmic
variable stars (e.g., Young \& Schneider 1980; Young, Schneider, \&
Shectman 1981; Horne \& Marsh 1986; Marsh 1988). It is now known that
the X-ray spectra of most Seyfert~1 galaxies possess broad, disk-like
iron K$\alpha$ lines (Tanaka et al. 1995; Mushotzky et al. 1995;
Nandra et al. 1997). Moreover, about 10\% of broad-line radio galaxies
show disk-like Balmer lines with twin peaks or twin shoulders
(Eracleous \& Halpern 1994; hereafter EH94). These disk-like Balmer
line profiles are approximately twice as broad as the ``normal'' broad
lines, and they are very well reproduced by models of gas rotating at
relativistic speeds in a Keplerian accretion disk. The relative
strengths of the narrow lines of these radio-loud ``disk-like''
emitters are similar to those found in LINERs (low-ionization nuclear
emission-line regions; Heckman 1980).

The association of broad, double-peaked Balmer lines (whatever their
origin) with objects displaying LINER-like narrow lines has been
underscored by the recent abrupt appearance of double-peaked Balmer
lines in three LINERs and LINER-like objects that did not previously
have them: NGC~1097 (Storchi-Bergmann, Baldwin, \& Wilson 1993);
Pictor~A (Halpern \& Eracleous 1994) and M81 (Bower et al. 1996).
While Pictor~A is a radio galaxy with similar characteristics to those
of the radio-loud ``disk-like'' emitters found by EH94, NGC~1097 and
M81 are spiral galaxies with low-luminosity LINER nuclei and much
weaker radio emission.  Nevertheless, NGC~1097 and M81 show
similarities with the more luminous double-peaked emitters found by
EH94, as both have nuclear non-thermal radio sources with a
synchrotron self-absorbed spectrum, characteristic of compact jets
(Hummel, van der Hulst, \& Keel 1987; Bartel et al. 1982). Moreover,
NGC~1097 is famous for its optical ``jets'' (Wolstencroft \& Zealey
1975), while the nucleus of M81 harbors an elongated radio structure
found by VLBI mapping (Bietenholz et al. 1996). The nuclei of NGC~1097
and M81 are also hard X-ray sources (Iyomoto et al. 1996; Ishisaki et
al. 1996).

In the case of NGC~1097, Eracleous et al. (1995) were able to
reproduce the shape of the double-peaked profile with a model of an
elliptical accretion ring. An elliptical, rather than circular, ring
was required in order to reproduce a double-peaked line with the red
peak stronger than the blue one. In a subsequent paper,
Storchi-Bergmann et al. (1995) presented the results of monitoring the
NGC~1097 H$\alpha$ profile: after approximately two years, the profile
had become more symmetric, with the two peaks showing similar
heights. This change was interpreted in the context of the elliptical
accretion ring model as a result of relativistic precession of the
ring (advance of the pericenter), which yielded an estimate of the
black hole mass of about $10^6$ M$_\odot$.  Here we present a new
spectrum showing that the profile has continued to vary, in a manner
consistent with an origin in a precessing accretion disk. We do,
however, find it necessary to refine the original model of
Storchi-Bergmann et al. (1995) so that it can account
self-consistently for all of the available data.

\topinsert
\hbox{\vsize=5 truein
\vbox to 5 truein {\hsize=3 truein
\centerline{\psfig{figure=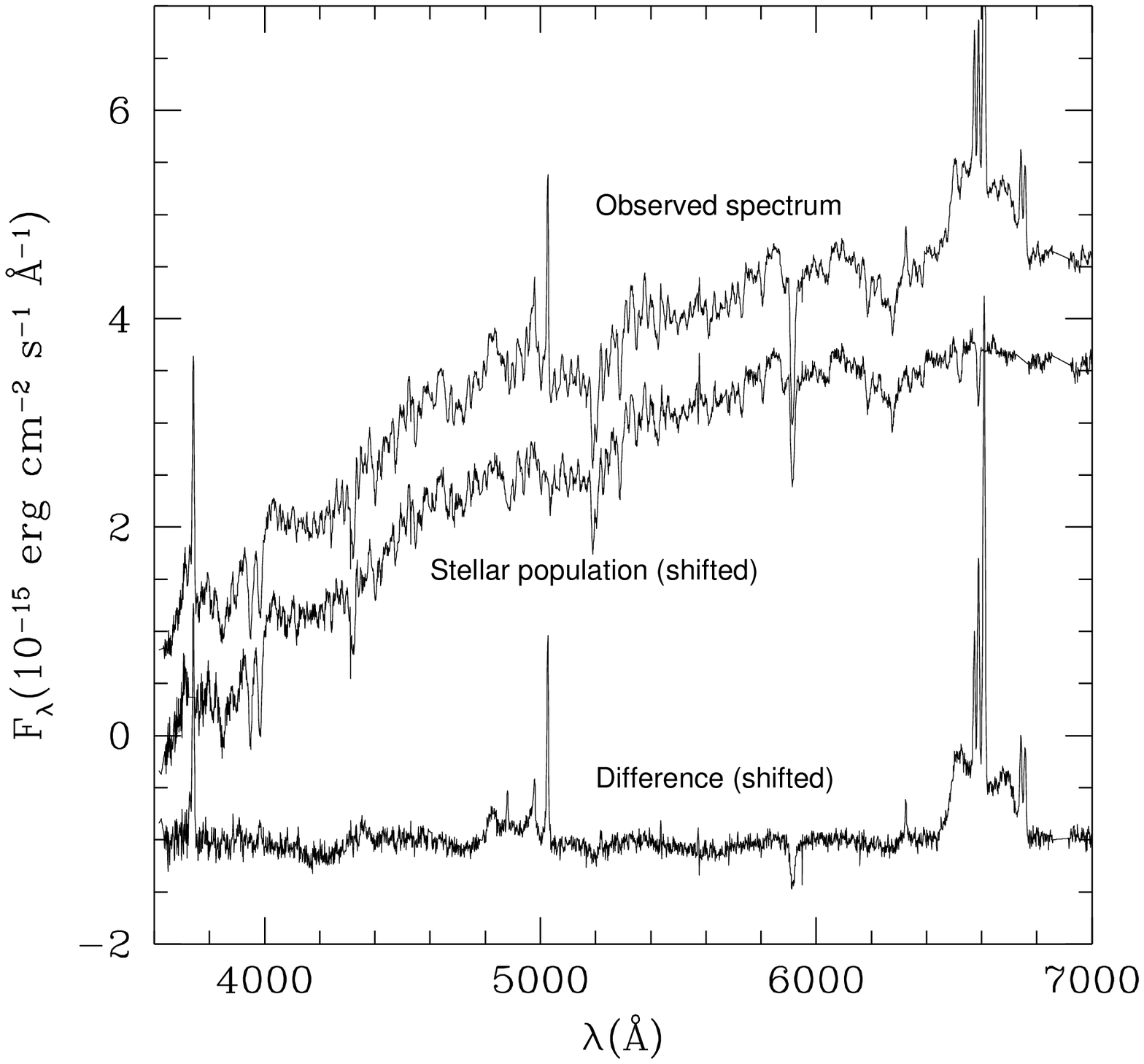,width=3.3in,rheight=3.2in,rwidth=3.5in}}
\medskip\noindent{\bf Figure 1 --}
The observed nuclear spectrum ({\it top}), the adopted stellar
population spectrum ({\it middle}) and the difference between the two
({\it bottom}), within an 2\sec$\times$4\sec\ aperture. The two last
spectra have been shifted for clarity.
\vfill} \hskip 0.5 truein
\vbox to 5 truein {\hsize=3 truein \vskip -0.35 truein
\centerline{\psfig{figure=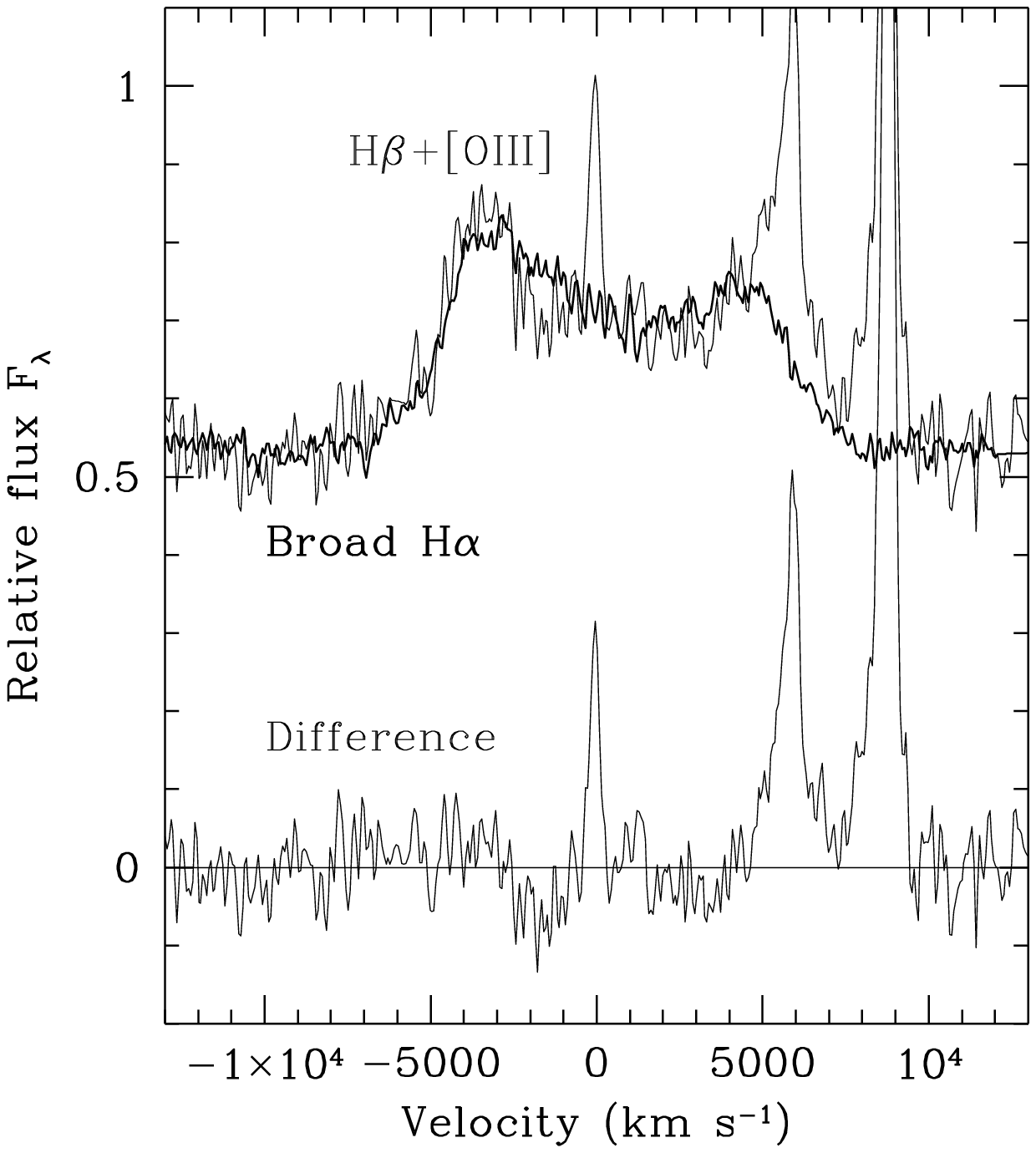,height=4in,rheight=3.5in,rwidth=4in}}
\medskip\noindent{\bf Figure 2 --}
{\it Top}: Comparison of the double-peaked H$\alpha$ profile
(heavy line), divided by 2.8 (see text), with the 
H$\beta$+[O~III] $\lambda\lambda$4959, 5007 profiles in velocity space.
{\it Bottom}: the difference (H$\beta$+[O~III]$)-$(H$\alpha/2.8)$. 
\vfill}}
\endinsert

\bigskip
\centerline{\bf 2. Observations and Results}
\medskip

A new long-slit optical spectrum of the NGC~1097 nucleus was obtained
using the Cassegrain Spectrograph and Loral 3K CCD detector at the 4~m
telescope of the Cerro Tololo Inter-American Observatory on 1996
January 24 UT.  The spectral range covered was $3500-7000$~\AA, at
4~\AA\ resolution. During the observation the 2$^{\prime\prime}$ slit
was oriented at position angle 83\deg. The zenith angle was small, and
therefore the effects of atmospheric dispersion are negligible. The
spectrum was reduced and calibrated using standard procedures and with
the help of the IRAF software package. In order to isolate the
emission from the gas, we subtracted the contribution of the stellar
population, which was obtained by averaging two spectra extracted from
approximately 5\sec\ E and 5\sec\ W of the nucleus. To match the slope
of the off-nuclear spectrum to the starlight spectrum of the nucleus
it was necessary to redden the former by $E(B-V)=0.12$~mag using a
Galactic reddening law (Seaton 1979).

Figure~1 shows the observed nuclear spectrum, together with the
starlight spectrum and the difference between the two. In the
starlight-subtracted spectrum, the double-peaked profile can be
readily seen, not only in H$\alpha$ but also in H$\beta$. The most
striking feature in the difference spectrum is the shape of the
double-peaked profiles: the blue peak is now stronger than the
red, the opposite asymmetry to that of the first observations of 1991
(Storchi-Bergmann et al. 1993). The flux in the double-peaked lines is
1.61$\pm$0.02$\times 10^{-13}$erg cm$^{-2}$ s$^{-1}$ for H$\alpha$ and
0.57$\pm0.02\times 10^{-13}$~erg cm$^{-2}$ s$^{-1}$ for H$\beta$. In
the difference spectrum the usual narrow lines are evident, as well as
the residual Na~I~D $\lambda$5892 interstellar absorption line, just
as in the spectra of Storchi-Bergmann et al. (1995). It may be argued
that a hint of broad Fe~II emission in the spectral regions around
4570 and 5300~\AA\ can also be seen, but a deeper exposure would be
necessary in order to confirm the presence of these features.

We show in Figure 2 a comparison of the H$\alpha$ and H$\beta$
profiles superposed in velocity space. The broad H$\alpha$ profile was
isolated by fitting and subtracting the narrow H$\alpha$,
[NII]~$\lambda\lambda$6548,~6584, and [SII]~$\lambda\lambda$6717,~6731
lines and then divided by 2.8, which is the observed ratio between the
integrated fluxes of the broad H$\alpha$ and H$\beta$ lines. We
conclude that, within the observational uncertainties, the two
profiles match each other very well. This is confirmed by the result
of subtracting the scaled broad H$\alpha$ from the H$\beta$+[OIII]
complex (bottom plot, Figure~2), which includes only the narrow
emission lines and residuals comparable to the noise in the difference
spectrum.  This result shows that in 1996 January there was neither
reddening of the region producing the double-peaked profiles, nor
differential reddening across the profile.

\bigskip
\centerline{\bf 3. Discussion}
\medskip
\centerline{\it 3.1 Comparison with Previous Observations}
\medskip

In Figure 3 we show the H$\alpha$ profile of the last observation
together with the two earlier spectra having similar spectral
resolution and signal-to-noise ratio. At the bottom of Figure~3 we
have plotted the H$\alpha$ profile of 1991 November 2, in the middle
that of 1994 January 4, and at the top that of 1996 January 24. These
spectra, spaced by approximately 2 years, represent three
characteristic snapshots in the continuous evolution of the H$\alpha$
profile over the course of our observations.

From the line fluxes measured in the starlight-subtracted 1996 spectrum, 
analysis of the profiles in Figure 3, and the previous measurements  
presented in Storchi-Bergmann et al. (1995), we conclude the following: 
\item{1.}
The integrated flux in the broad H$\alpha$ emission line, which decreased by
50\% from  1991 November to 1994 January, has increased  to 83\% 
of its original flux.
\item{2.}
The two peaks have kept the same velocities,
$-3300\pm$200~km~s$^{-1}$ for the blue peak and 3900$\pm$200 km~s$^{-1}$ 
for the red, relative to the narrow component of H$\alpha$. This implies 
that the H$\alpha$ line as a whole has not shifted appreciably. Thus, the
average velocity of the two peaks relative to the narrow H$\alpha$ is still
$300\pm 50$ km~s$^{-1}$. We note that the heliocentric velocity of the 
narrow emission lines is consistent, within measurement uncertainties,
with the heliocentric velocity derived from the stellar absorption lines
(the two velocities differ by less than 40~km~s$^{-1}$).
\item{3.}
The 1996 January profile of H$\alpha$ is almost a mirror image of the 1991
November profile.
\item{4.}
The detectable limit of the red wing relative to the narrow component
of H$\alpha$ is $8400\pm$500~km~s$^{-1}$, and that of the blue wing is
$-7200\pm$500~km~s$^{-1}$, except in 1991 November, when the blue wing
was more extended to about $-12500\pm1000$~km~s$^{-1}$.
\item{5.}
The H$\alpha$/H$\beta$ ratio, which increased from 3.2$\pm$0.5 in 1991
November to 4.2$\pm$0.7 in 1993 September, has now decreased to
2.8$\pm$0.3.

\noindent
In Figure~4 we show some of the above conclusions graphically by plotting
the variation of the broad H$\alpha$ flux and the H$\alpha$/H$\beta$ ratio 
over the course of our observations.

\topinsert
\hbox{\vsize=5.5 truein
\vbox to 5.5 truein {\hsize=3 truein
\centerline{\psfig{figure=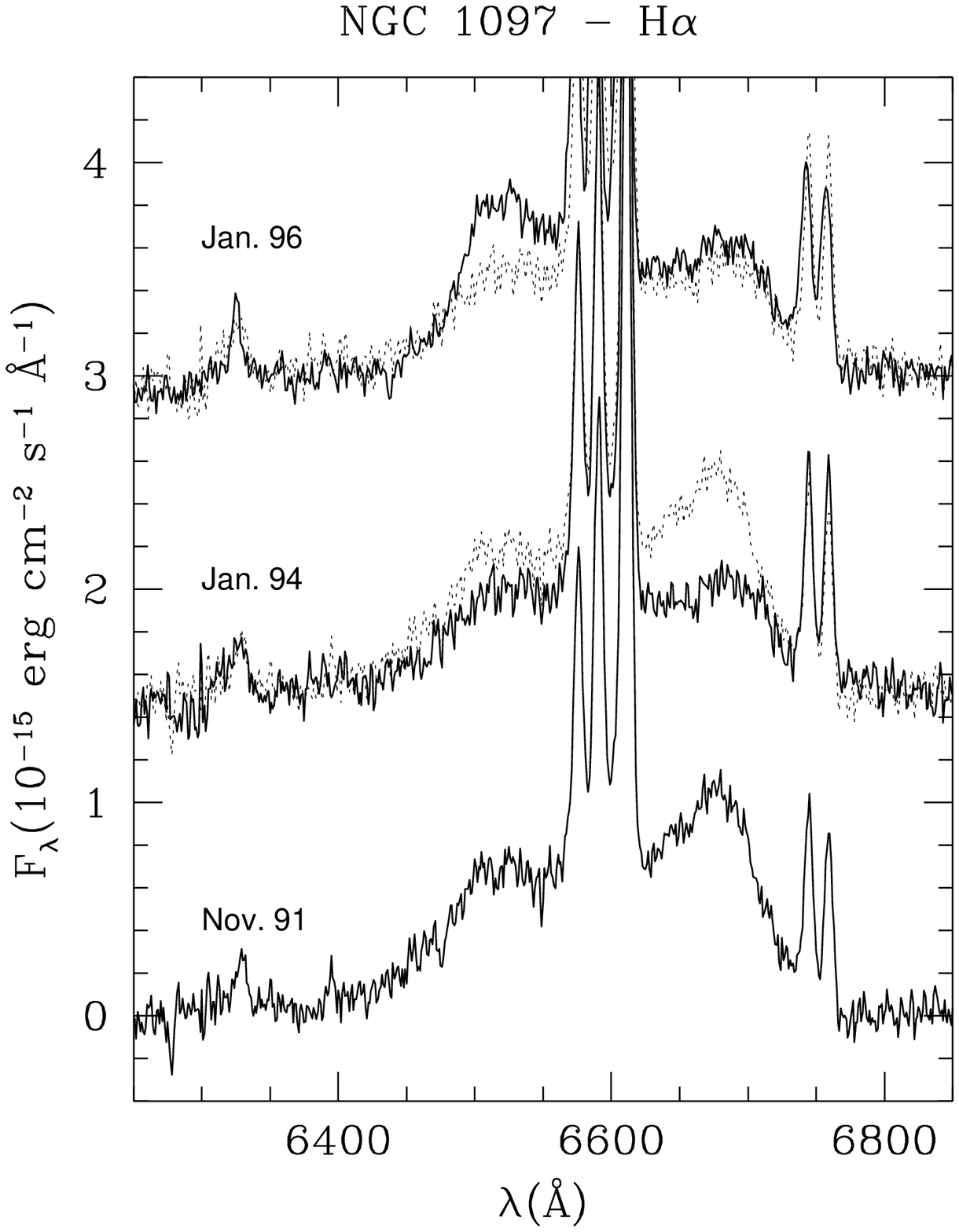,height=4in,rheight=4in,rwidth=4in}}
\medskip\noindent{\bf Figure 3 --}
The double-peaked H$\alpha$ profiles in three distinct epochs. Dashed
lines show the profile from the previous epoch, for comparison.
\vfill} \hskip 0.5 truein
\vbox to 5.5 truein {\hsize=3 truein \vskip -0.4 truein
\centerline{\psfig{figure=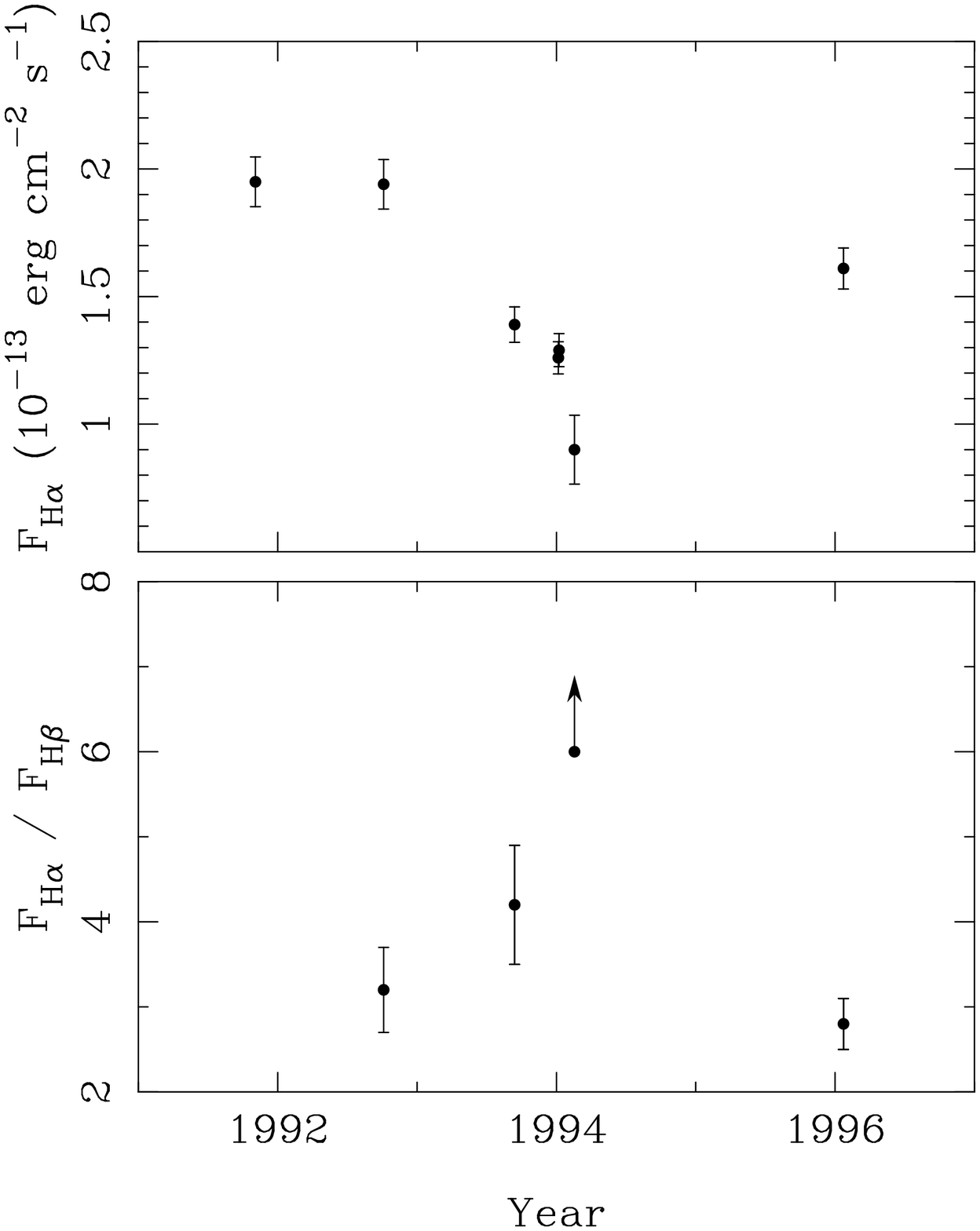,height=4.5in,rheight=4.0in,rwidth=3.3in}}
\vfill
\medskip\noindent{\bf Figure 4 --} 
The variation of the broad H$\alpha$ flux ({\it top}) and the 
H$\alpha$/H$\beta$ ratio ({\it bottom}) over the course of the NGC~1097 
observations.
\vfill}}
\endinsert

\medskip
\centerline{\it 3.2. Line Emission from an Eccentric Accretion Ring}
\medskip

The new spectrum of NGC~1097 presented here shows that the strength of
the broad, double-peaked Balmer lines has recovered from the decline
it was undergoing in 1994. The line profiles have been changing
continuously since the first detection of the broad lines in 1991
November (see Storchi-Bergmann et al. 1993, 1995). The red peak of
H$\alpha$, which was initially stronger than the blue peak, has been
steadily declining in strength. In the most recent spectrum the blue
peak has become stronger than the red one (see Figures~1 and 3). This
suggests very strongly that the observed line profile variations are
the result of a continuous process, which seems to operate
independently of the variations in the integrated flux of the Balmer
lines.

Storchi-Bergmann et al. (1995) considered a number of possible
scenarios to account for the profiles of the double-peaked Balmer
lines and their variability, favoring in the end a model of an
eccentric accretion ring (Eracleous et al. 1995). Such an eccentric
ring is expected to result from the tidal disruption of a star by a
supermassive black hole in the galactic nucleus (e.g., Rees 1988). The
tidal disruption hypothesis is appealing because it can explain the
abrupt appearance of the broad Balmer lines, and also because the
formation and subsequent precession of an eccentric disk can explain
the profiles of the broad lines and their variability. Storchi-Bergmann 
et al. (1995) also argued that the variation of the {\it profiles} of the 
broad lines and the variation of the integrated line fluxes result
from independent processes, which is supported by the new
spectrum. Two plausible explanations were put forward for the decline
of the integrated flux of the broad Balmer lines: (a) the flux of
ionizing radiation was itself declining, presumably as a result of a
decreasing accretion rate, and (b) an obscuring cloud was crossing the
line of sight to the line-emitting ring. (This cloud could be part of
the interstellar medium in the immediate vicinity of the nucleus or a
portion of the post-disruption debris that was able to expand
adiabatically and reach a large height out of the plane of its
original trajectory.)  Both of these hypotheses could account for the
steepening of the Balmer decrement that accompanied the decline in the
broad line flux (see Figure~4), but it was predicted that if the
latter hypothesis is true, the broad lines should recover after the
obscuring material left the line of sight. This is indeed what the new
data show.

The elliptical ring model for the line profile and the tidal
disruption scenario for the appearance of the lines were not only
successful in explaining the observations, but their prediction for
the evolution of the profile has now received strong support. In
addition to accretion-disk emission, two alternative scenarios are
considered for the origin of broad, double-peaked emission lines: the
bipolar outflow scenario (Zheng, Binette, \& Sulentic 1990) and the
binary black hole hypothesis (Gaskell 1983). Independently of the
variability properties of the line profiles, the binary black hole
hypothesis appears unlikely because it requires that the supermassive
binary in NGC~1097 would have to be of an unlikely type.  The
separation of the two peaks in the Balmer line profiles implies that
the two black holes should have roughly equal masses.  If such a
binary is to form from the merger of two parent galaxies, then the two
parents should also have roughly equal masses. In all cases where the
masses of black holes in the centers of active and non-active galaxies
have been determined by stellar or gas kinematics, the mass of the
black hole is approximately proportional to the mass of the bulge of
the parent galaxy (Kormendy et al. 1997). Therefore, the result of the
merger would not be spiral galaxy like NGC~1097. Models in which the
double-peaked lines originate in the oppositely directed parts of a
bipolar outflow are purely kinematic and thus notoriously difficult to
constrain. These models can possibly accommodate the observed
profile variability although they cannot explain it in terms of a
specific physical mechanism. Therefore we do not consider these
scenarios further and concentrate instead on accretion disk models. In
the remainder of this section we consider the eccentric ring model in
detail, while in the next section we discuss an alternative phenomenon
associated with an accretion disk which may produce a similar behavior
of the line profiles.

The specific model proposed by Storchi-Bergmann et al. (1995) involved
an eccentric ring with a mean pericenter distance of 2600~$r_{\rm g}$
(where $r_{\rm g}\equiv GM/c^2$ is the gravitational radius, with $M$
the mass of the black hole, $G$ the gravitational constant, and $c$
the speed of light), and a constant eccentricity of 0.5. In this
scenario, the precession of the ring, resulting from the general
relativistic advance of the pericenter, was responsible for the smooth
variation of the H$\alpha$ line profile. The corresponding model
profiles were in qualitative agreement with the observations. The
inferred precession rate led to an estimate of the mass of the central
black hole of approximately $10^6$~M$_{\odot}$. We find it necessary
to revise this model somewhat for two reasons. First, the direction of
precession of the ring that was used to compute the model profiles was
the opposite from what general relativistic advance of the pericenter
would produce, and was thus physically incorrect. The correct
direction of precession produces a change in the line profiles in the
opposite sense to what is observed. Second, regardless of this
error, a model involving an elliptical ring with a constant
eccentricity is not able to fit all of the observed spectra in
detail: the new spectrum presented here poses the most severe
difficulty.

The most important revision that we propose is to make the
eccentricity of the ring a function of radius (more precisely, of
pericenter distance). We construct a model in which the inner annulus
of the ring is circular and the eccentricity grows linearly with
radius up to a maximum value of 0.45. The plane of the ring is
inclined at 34$^{\circ}$ to the line of sight, and the line-emitting
part of the ring is bounded by annuli with pericenter distances of
1300~$r_{\rm g}$ and 1600~$r_{\rm g}$. The ring is circular up to
1400~$r_{\rm g}$. At this radius it develops an eccentricity which
increases linearly to its maximum value at the outer radius. An
eccentricity that increases with radius is motivated by the results
of the simulations of Syer \& Clarke (1992), which follow the
evolution of the post-disruption debris from the tidal encounter of a
star with a black hole. When the debris settles into a well-formed
disk after about a viscous time, the inner parts have circularized as
a result of differential precession, while the eccentricity of the
outer particle orbits increases approximately linearly with
radius. The model we are considering is purely kinematic and it uses a
parametric description of the accretion flow. As such, it may not
conform well to the results of detailed hydrodynamical
calculations. However, it allows us to draw conclusions about the
general kinematic properties of the line-emitting gas. The modification
that we have introduced adds one more free parameter to the disk model 
discussed by Storchi-Bergmann et al (1995), which involved seven free 
parameters. One of the features of the refined model is that the predicted 
velocity shifts of the twin peaks resulting from the precession
of the ring are small. This is a result of the combination of the steep
emissivity law and the low eccentricity of the inner annuli of 
the disk where the emission is most intense.

In Figure~5 we show the detailed fits of the refined elliptical ring
model to the observed line profiles. We have chosen the three
high-resolution spectra from our collection as examples; these also
represent the beginning, middle, and end of the time sequence that we
have observed so far. In the sequence of models presented here we have
allowed only for a precession of the elliptical ring, holding all
other parameters fixed. In the 4.2 years covered by the spectra of
Figure~5, the elliptical ring precesses by 60$^{\circ}$ (i.e., the
azimuthal angle of the apocenter increases smoothly from 320\deg\ to
20\deg), which implies a precession period of 25 years, and in turn a
central black hole mass of $9\times10^5$~M$_{\odot}$. We have arrived
at this model by trial and error, judging the goodness of the fit by
eye. At each individual epoch the best-fitting model is not
unique. For example, the H$\alpha$ model in 1991 November can be
fitted equally well by the model presented by Eracleous et al.  (1995)
and by the model presented here. However, the information on the
evolution of the profile with time restricts the possible models
considerably.  The inclination and the inner and outer radii of the
ring can be constrained fairly well (within 10\%) by fitting the wings
of the line profile. With these parameters held fixed, the orientation
and eccentricity of the disk can be determined (to a similar
uncertainty) by the requirement that the evolution of the profile is
reproduced.

\topinsert
\hbox{\vsize=4.7 truein \hskip -0.2 truein
\psfig{figure=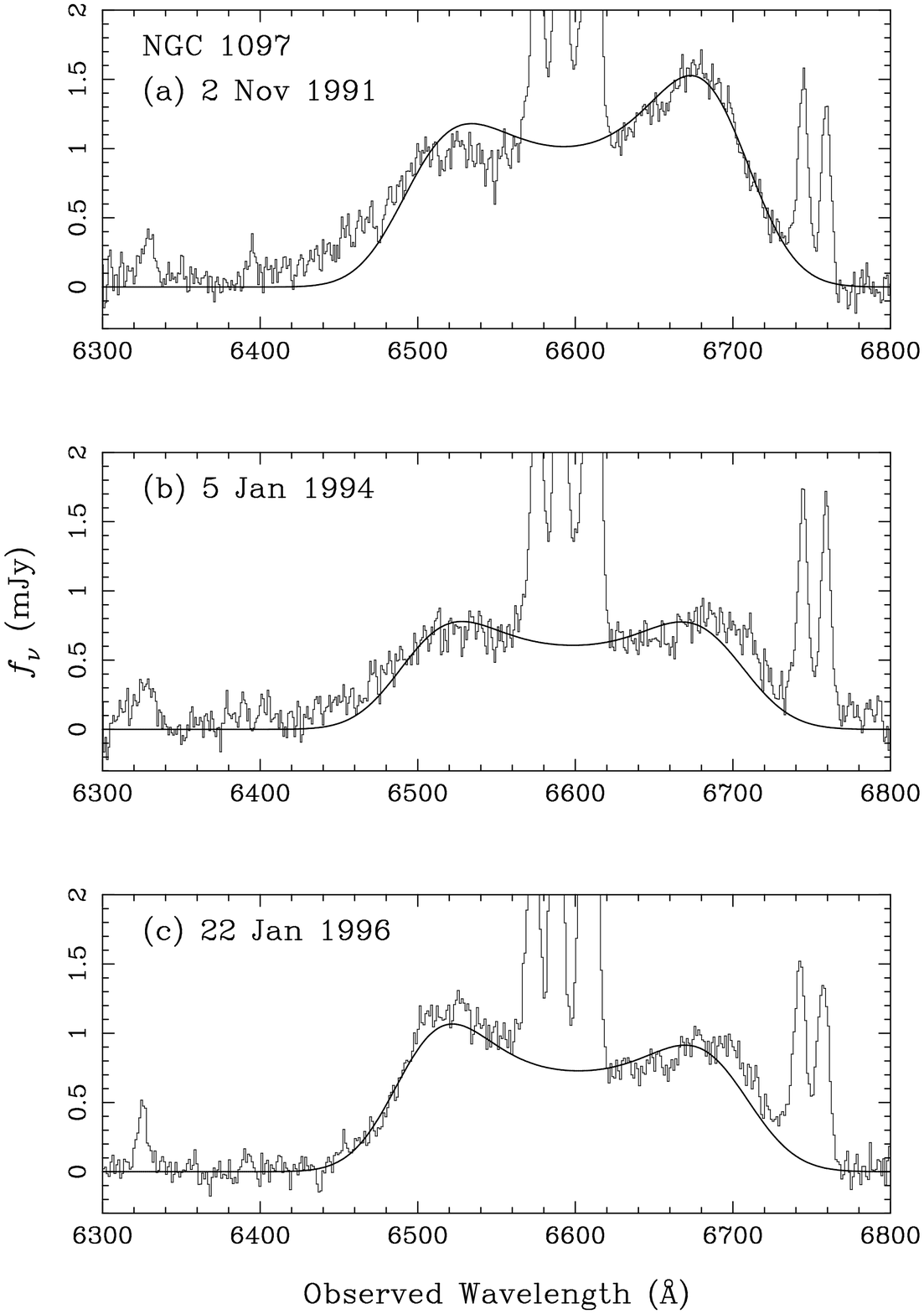,height=4in,rheight=4.5in,rwidth=3in}
\hskip 0.3 truein
\vbox to 4.7 truein {\hsize=3 truein
\medskip\noindent{\bf Figure 5 --}
Detailed fits of the elliptical ring model to the three high resolution 
H$\alpha$ profiles of NGC~1097. In all cases the continuum has been subtracted 
using an off-nuclear starlight spectrum (see text for details). The line
emitting ring has a circular inner annulus, followed by elliptical outer
annuli whose eccentricity increases linearly outwards up to a maximum value.
The normal to the plane of the ring is inclined by 34\deg to the line of sight.
The inner circular annulus spans the radial range between 
1300~$r_{\rm g}$ and 1400~$r_{\rm g}$, and the pericenter distance of the outer
annulus  is 1600~$r_{\rm g}$. The maximum  (outer) eccentricity is 0.45. In (a)
the observer is located at an azimuth of 320\deg (see definition in Eracleous et
al. 1995).  In (b) and (c) the ring is allowed to precess in steps of 30\deg.
\vfill}}
\endinsert

\medskip
\centerline{\it 3.3. Line Emission from a Precessing, Warped Disk}
\medskip

\def\gs{\lower 2pt \hbox{$\;\scriptscriptstyle \buildrel>\over\sim\;$}} 
\def\ls{\lower 2pt \hbox{$\;\scriptscriptstyle \buildrel<\over\sim\;$}} 

The eccentric ring scenario discussed above is successful in
accounting for all of the observed features of the Balmer lines,
including their profiles and their variability. It is still
worthwhile, however, to consider alternative phenomena in the disk
which could produce (qualitatively at least) a similar behavior. More
specifically, we discuss a mechanism for producing the smooth and
continuous line profile variation shown by the observations. We
envision a scenario in which an accretion disk forms abruptly by the
tidal disruption of a star by a supermassive black hole, much like
what has been discussed above. However, we ascribe the variation of
the line profile to a perturbation in the brightness of the disk (a bright
spot) which revolves smoothly around the center. A similar scenario
has been used to explain the variation of the double-peaked H$\alpha$
emission line of Arp~102B by Newman et al. (1997).  We propose that
the revolving bright spot is a precessing warp in the disk, induced by
irradiation from the center, and we examine below the qualitative
features of this hypothesis.

In a recent work, Pringle (1996) demonstrated that accretion disks
irradiated by a central source can become unstable to warping
(see also Petterson 1977, and Iping \& Petterson 1990). A warped disk
is expected to precess. It has been shown that such
irradiation-induced warps can occur in X-ray binaries, in the masing
disk of NGC~4258, and in ``point-symmetric'' planetary nebulae
(Maloney, Begelman, \& Pringle 1996; Livio \& Pringle 1996;
Southwell, Livio, \& Pringle 1997). Here we examine whether an
accretion disk in the central engine of NGC~1097 could be subject to
this instability, and we estimate the expected precession time scale.

The condition for the instability can be expressed as (Pringle 1996)
$$
{R\over r_{\rm g}} \gs\; \xi\; {\eta^2\over \epsilon^2}\; ,
\eqno{(1)}
$$
where $R$ is the radial distance in the disk, $r_{\rm g}$ is the
gravitational radius defined earlier, $\eta=\nu_1/\nu_2$ is the ratio
of the $(R,z)$ and $(R,\phi)$ viscosities, $\epsilon=L/{\dot m} c^2$
is the accretion efficiency (with $L$ the accretion luminosity and
$\dot m$ the mass accretion rate), and $\xi$ is a dimensionless
parameter. Analytic estimates give $\xi\approx 16\,\pi^2$, while
numerical results indicate that $\xi\gs 8$. Taking $\eta=1$ and
assuming a typical accretion efficiency of $\epsilon\approx 0.1$, we
find that the disk can become unstable to warping for $R$ around 800
to 1600~$r_{\rm g}$. This means that the disk in NGC~1097, which would
have a radius between 1000~$r_{\rm g}$ and 2000~$r_{\rm g}$, {\it
could} become unstable.

The precession time scale of the disk is given by (Pringle 1996)
$$
\tau_{\rm prec} \approx {12\,\pi\, c\,\Sigma\, R^3\,\Omega \over L}\; ,
\eqno{(2)}
$$
where $\Sigma$ is the surface density and $\Omega$ is the orbital
angular velocity. We now proceed to estimate this time scale for the
particular case of NGC~1097. We assume the mass of the central black
hole to be $M\approx 10^6$~M$_{\odot}$.  A mass of this order is
favored in tidal disruption scenarios for the formation of the disk
since a black hole of this mass would readily disrupt a star before
accreting it (Rees 1988). The X-ray luminosity of the nucleus was
found to be $L_{\rm X} \approx 2\times 10^{41}$ erg s$^{-1}$, from
observations with the $ROSAT$ High-Resolution Imager (P\'erez-Olea \&
Colina 1996). The mass of the H$\alpha$ emitting gas was estimated by
Storchi-Bergmann et al. (1995) to be
$$
M_{\rm H\alpha} \approx 0.24\; \left({n\over 10^8~{\rm cm}^{-3}}\right)^{-1}
\; \left({H_0\over 75~{\rm km~s^{-1}~Mpc^{-1}}}\right)^{-2}\;{\rm M_{\odot}}\; ,
\eqno{(3)}
$$
where $n$ is the electron density and $H_0$ is the Hubble constant. If
the density of the line emitting gas is in the range expected for a
standard broad-line region (i.e., of order ${\rm 10^{10}~cm^{-3}}$ or
greater), the observed H$\alpha$/H$\beta$ ratio of about 3, suggests a
value of order ${\rm 10^{12}~cm^{-3}}$ (Rees, Netzer, \& Ferland
1989). Of course, gas with a very low density (on the order of ${\rm
10^3~cm^{-3}}$) can also produce an H$\alpha$/H$\beta$ of about 3 but
if this was the case, the forbidden lines would also be broad and
double peaked (the Balmer lines are the only broad, double-peaked
lines observed). Thus the mass of ionized gas in the visible portion of
the disk is of order $M_{\rm ring}\approx 10^{-4}~{\rm M_{\odot}}$. We
emphasize that this estimate of the mass refers to the portion of
the elliptical ring that emits the H$\alpha$ line. A larger reservoir
of matter (e.g., $10^{-3} - 10^{-2}$~M$_{\odot}$) may exist exterior
to this ring, where the disk is thicker and denser (see, for example,
Collin-Souffrin \& Dumont 1990).  For parameters typical of NGC~1097,
equation (2) therefore gives
$$
\tau_{\rm prec} \sim 8.8\; 
\left({M_{\rm ring}\over 10^{-4}~{\rm M_{\odot}}}\right)
\left({M\over 10^6~{\rm M_{\odot}}}\right)^{1/2}
\left({R\over 1500\; r_{\rm g}}\right)^{-1/2}
\left({L\over 2\times 10^{41}~{\rm erg~s^{-1}}}\right)^{-1}
\; {\rm yr.} \eqno{(4)}
$$
This precession time scale is of the right order to explain the observed
changes in the profile. A detailed comparison of this model with the
observational data would involve detailed modeling of the emission from a 
warped disk, which is beyond the scope of the present work.

\medskip
\centerline{\bf 4. Conclusions and Speculations}
\medskip
The behavior of the double-peaked Balmer lines of NGC~1097 is so far
consistent with the scenario attributing their origin to an accretion
disk which has formed abruptly from the tidal disruption of a star by
a black hole. There are a number of different phenomena in the disk
that can cause it to be non-axisymmetric and, in turn, result in
asymmetries and time variability of the profiles of the lines that it
emits. The eccentric precessing disk model discussed above can fit the
line profiles quite well, and leads to a reasonable dynamical estimate
of the mass of the central black hole. The warped disk model is also
plausible (at least qualitatively) since it can reproduce the observed
variability time scale.  The fact that the emission line profiles have
been varying smoothly and continuously over the past 4 years
reinforces our proposal that they originate in a single structure in
the accretion flow, such as the accretion disk.

The two disk scenarios that we have discussed above can be used to
make predictions (albeit uncertain) about the future evolution of the
line profiles.  In the case of the irradiation-induced warp, we expect
that the pattern of variability that has been observed so far will
continue for some time, although eventually (when shadowing starts to
play an important role in the disk) the behavior could become chaotic
(Livio \& Pringle 1997). This is expected to happen when the
inclination angle of some part of the disk changes by more than
90$^\circ$. In the case of the eccentric precessing disk model, we
also expect this pattern of variability to continue, and to be
periodic, at least in principle. However, the evolution of the profile
shape in this model depends sensitively on the relative time scales
for precession and circularization. The latter time scale is expected
to be longer than the former (see the estimate of Eracleous et
al. 1995), but this question cannot be answered with confidence
without detailed numerical calculations.

Whatever the evolution of the profiles of the double-peaked lines, it
is worthy of study. An alternative model that is not discussed in
detail here but deserves attention nonetheless is one in which the
abrupt appearance of the double-peaked Balmer lines is the result of
the sudden illumination of a pre-existing accretion disk by a
transient source of ionizing radiation. We speculate that such a
transient ionizing source could be associated either with the
formation of an elevated structure in the inner accretion disk (e.g.,
an ion-supported torus; Rees et al. 1982) or with an outburst
analogous to a dwarf nova event in the disk.  In this context the
variations in the integrated flux of the line would be attributed to
variations in the intensity of the ionizing source, and the variations
of the line profile could be the result of a bright spot in the disk,
or a warp induced by the intense illumination.  In order to make
further progress, the models proposed here should be tested by future
observations. The variability of the line profiles is an important
aspect of the proposed scenarios, and offers a way of testing them by
continued spectroscopic monitoring.  The phenomenon of the abrupt
appearance of double-peaked emission lines in LINERs and LINER-like
objects is not confined to NGC~1097, but extends to a few other
objects as well (M81, Pictor~A; see \S~1). NGC~1097 is the most
spectacular performer among these objects, and therefore deserves
frequent and careful spectroscopic monitoring. We would not like to
miss the next episode of this drama!

\bigskip \noindent
T.S.-B. acknowledges fruitful discussions with Luc Binette and Charles
Bonatto, and partial support from the Brazilian institutions CNPq,
CAPES, and FAPERGS.  We also thank the referee, R. Antonucci, for
useful comments.  M.E. acknowledges support from NASA through the
Hubble Fellowship grant HF-01068.01-94A awarded by the the Space
Telescope Science Institute (STScI), which is operated for NASA by the
Association of Universities for Research in Astronomy, Inc., under
contract NAS~5-26255. M.L. and A.S.W.acknowledge support from NASA grants
NAGW-2678 and NAGW-4700, respectively and A.V.F. acknowledges support
from STScI grant AR-05792.01-94A.

\bigskip
\centerline{\bf References}
\medskip

\def\ref#1{{\par\noindent \hangindent=3em\hangafter=1 #1\par}}
\def\ditto{\vrule width1.2cm height3pt depth-2.5pt}

\ref  {Antonucci, R. R. J. 1992, in Testing the AGN Paradigm,
      ed. S. S. Holt, S. G. Neff, \& C. M. Urry, (New York: AIP), 486}
\ref {Antonucci, R. R. J., Kinney, A. L., \& Ford, H. C. 1989 ApJ, 342, 64}
\ref {Bartel, N., et al. 1982, ApJ, 262, 556}
\ref {Bietenholz, M. F., et al. 1996, ApJ, 457, 604}
\ref {Bower, G. A., Wilson, A. S., Heckman, T. M., \& Richstone, D. O. 1996, 
      AJ, 111, 1901}
\ref {Collin-Souffrin, S., \& Dumont, A. M. 1990, A\&A, 229, 292}
\ref {Eracleous, M., \& Halpern, J. P. 1994, ApJS, 90, 1 (EH94)}
\ref {Eracleous, M., Livio, M., Halpern, J. P., \& Storchi-Bergmann, T. 1995,
      ApJ, 438, 610}
\ref {Gaskell,  C. M. 1983, in Proc. 24th Li\`ege Int. Astrophys. Colloq.
       (Cointe-Ougree: Univ. Li\`ege), 473}
\ref {Halpern, J. P., \& Eracleous, M. 1994, ApJ, 433, L17}
\ref {Heckman, T. M., 1980, A\&A, 87, 152}
\ref {Horne, K., \& Marsh, T. R. 1986, MNRAS, 218, 761}
\ref {Hummel, E., van der Hulst, J. M., \& Keel, W. C. 1987, A\&A, 172, 32}
\ref {Iping, R. C., \& Petterson, J. A. 1990, A\&A, 239, 221}
\ref {Ishisaki, Y., et al. 1996, PASJ, 48, 237}
\ref {Iyomoto, N., Makishima, K., Fukazawa, Y., Tashiro, M.,
      Ishisaki, Y., Nakai, N., \& Taniguchi, Y. 1996, PASJ, 48, 231}
\ref {Koratkar, A., Kinney, A. L., \& Bohlin, R. C. 1992, ApJ, 400, 435}
\ref {Koratkar, A., Antonucci, R. R. J., Goodrich, R. W., Bushouse, H., \& 
      Kinney, A. L. 1995, ApJ, 450, 501}
\ref{Kormendy, J. et al. 1997, ApJ, in press}
\ref {Livio, M. \& Pringle, J. E. 1996, ApJ, 465, L55}
\ref {\ditto\ . 1997, ApJ, submitted}
\ref {Maloney, P. R., Begelman, M. C., \& Pringle, J. E. 1996, ApJ, 472, 582}
\ref {Marsh, T. R. 1988, MNRAS, 231, 1117}
\ref {Mushotzky, R. F. et al., 1995, MNRAS, 272, P9}
\ref {Nandra, K., George, I. M., Mushotzky, R. F., Turner, T. J., \&
      Yaqoob, T. 1997, ApJ, in press}
\ref {Newman, J. A., Eracleous, M., Halpern, J. P., \& Filippenko, A. V. 1997,
      ApJ, in press}
\ref {P\'erez-Olea, D. E., \& Colina L. 1996, ApJ, 468, 191}
\ref {Petterson, J. A. 1977, ApJ, 216, 827}
\ref {Pringle, J. E. 1996, MNRAS, 281, 357}
\ref {Rees, M. J. 1988, Nature, 333, 523}
\ref {\ditto\ . 1990, Science, 247, 817}
\ref {Rees, M. J., Begelman, M. C., Blandford, R. D., \& Phinney, E. S. 
      1982, Nature, 295, 17}
\ref {Rees, M. J., Netzer, H., \& Ferland, G. J. 1989, ApJ, 347, 640}
\ref {Seaton, M. J. 1979, MNRAS, 187, 73P}
\ref {Southwell, K. A., Livio, M., \& Pringle, J. E. 1997, ApJ, in press}
\ref {Storchi-Bergmann, T., Baldwin, J. A., \& Wilson, A. S. 1993, 
      ApJ, 410, L11}
\ref {Storchi-Bergmann, T., Eracleous, M., Livio, M., Wilson, A. S.,
      Filippenko, A. V., \& Halpern, J. P. 1995, ApJ, 443, 617}
\ref {Syer, D., \& Clarke, C. J. 1992, MNRAS, 255, 92}
\ref {Tanaka, Y., et al. 1995, Nature, 375, 659}
\ref {Wolstencroft, R. D., \& Zealey, W. J. 1975, MNRAS, 173, 51P}
\ref {Young, P., \& Schneider, D. P. 1980 ApJ, 238, 955}
\ref {Young, P., Schneider, D. P., \& Shectman, S. A. 1981 ApJ, 245, 1035}
\ref {Zheng, W, Binette, L., \& Sulentic, J. W. 1990, ApJ, 365, 115}

\end